\documentclass[11pt,a4paper]{article}
\usepackage{amsmath,amssymb,amsfonts,a4wide,graphicx,bm,times}
\usepackage{mcite}
\usepackage{hyperref}
\hypersetup{
    colorlinks,
    citecolor=green,
    filecolor=green,
    linkcolor=blue,
    urlcolor=green
}

\makeatletter \let\old@startsection=\@startsection
\renewcommand{\@startsection}[6]
{\old@startsection{#1}{#2}{#3}{#4}{#5}{#6\mathversion{bold}}}
\makeatother

\makeatletter

\@addtoreset{equation}{section}
\makeatother
\let\refOld\ref
\newcommand{\superp}[2]{\genfrac{}{}{0pt}{}{#1}{#2}}

 \def\d{\delta}

 \def\p{\partial}
 
 \def\a{\alpha}
 
 \def\g{\gamma}
 \def\d{\delta}
 \def\e{\varepsilon}

 \def\k{\kappa}
 \def\l{\lambda}

 \def\r{\rho}
 
 \def\s{\sigma}
 \def\t{\tau}

 \def\G{\Gamma}
 \def\D{\Delta}

 \def\L{\Lambda}
 \def\O{\Omega}

\def\CF{{\mathcal{F}}}

\def\CN{{\mathcal{N}}}

\def\CS{{\mathcal{S}}}

\def\CZ{{\mathcal{Z}}}
\def\la{\left\langle}
\def\ra{\right\rangle}

\def\hf{\dfrac{1}{2}}

\def\vphi{\varphi}

\def\CF{\mathcal{F}}

\def\CFs{\CF_\text{short}}
\def\Ys{Y_\text{short}}
\def\YsT{Y_{T}}

\begin{document}
\begin{titlepage}
\renewcommand{\thefootnote}{\fnsymbol{footnote}}
\vspace*{-2cm}
\begin{flushright}
\end{flushright}

\vspace*{1cm}
    \begin{Large}
%    \begin{bf}
       \begin{center}
         {\huge Finite $\e_2$-corrections to the $\mathcal{N}=2$ SYM prepotential}
       \end{center}
%    \end{bf}   
    \end{Large}
\vspace{0.7cm}

\begin{center}
Jean-Emile Bourgine, Davide Fioravanti\footnote{e-mail addresses: bourgine@bo.infn.it, fioravanti@bo.infn.it}\\
      
\vspace{0.7cm}   
{\em Sezione INFN di Bologna, Dipartimento di Fisica e Astronomia,\\
Universit\`a di Bologna} \\
{\em Via Irnerio 46, Bologna, Italy}\\   
              
\end{center}

\vspace{0.7cm}

\begin{abstract}
\noindent
We derive the first $\e_2$-correction to the instanton partition functions of $\mathcal{N}=2$ Super Yang-Mills (SYM) in four dimensions in the Nekrasov-Shatashvili limit $\e_2\rightarrow 0$. In the latter we recall the emergence of the famous Thermodynamic Bethe Ansatz-like equation which has been found by Mayer expansion techniques.  Here we combine efficiently these to field theory arguments. In a nutshell, we find natural and resolutive the introduction of a new operator $\nabla$ that distinguishes the singularities within and outside the integration contour of the partition function.
\vspace{0.5cm}
%\noindent\textbf{Keywords:} Mayer Expansion, Thermodynamical Bethe Ansatz, $\mathcal{N}=2$ SUSY Gauge Theory, Instantons\\
\end{abstract}

\vfill

\end{titlepage}
\vfil\eject

\setcounter{footnote}{0}

% \vspace*{-2cm}
% \begin{flushright}
% \jobname .pdf\\ \today
% \end{flushright}
% 
% \begin{center}
% \huge Subleading $\e_2$-corrections to the Nekrasov-Shatashvili prepotential.
% \end{center}
% 
% \hspace{2cm}

\section{Introduction}
The Omega background was first introduced in $\mathcal{N}=2$ SUSY gauge theories to regularise the infinite volume of $\mathbb{R}^4$ in the computation of instanton contributions to the partition function by localisation \cite{Nekrasov2003}. More recently, it has also proven to be a formidable way to preserve integrability of these theories upon deformation. In fact, this background is characterised by two equivariant deformation parameters $\e_1$ and $\e_2$ associated to the breaking of the Lorentz invariant four dimensional space into $\mathbb{C}\times\mathbb{C}$, but still the Nekrasov instanton partition function exhibits an integrable structure in the form of covariance under the Spherical Hecke central (SHc) algebra \cite{Schiffmann2012,Kanno2013} (which is formally equivalent to a $W_\infty$ algebra). The presence of this algebra shed light on the conjecture by \cite{Alday2009,Wyllard2009} of a duality between these four dimensional theories and the family of Toda conformal field theories in two dimensions.\footnote{The conjecture has been partially proved either using the basis of AFLT states \cite{Alba2010}, or a set of generalized Jack polynomials \cite{Morozov2013,Mironov2013}.}  Moreover, the SHc algebra is closely related to a tensorial version of the integrable Calogero-Moser Hamiltonian and led to the construction of one of the most basic objects of quantum integrable theories, namely a (instanton) $R$-matrix \cite{Maulik2012}.

In the Nekrasov-Shatashvili (NS) limit $\e_2\rightarrow 0$, the theory possesses only one (non-zero) equivarant parameter $\e_1$ and, besides, its 2D underlying integrable structure is better understood \cite{Nekrasov2009}. The latter defines, indeed, a quantisation of a Hitchin system associated to the Seiberg-Witten curve (see \cite{Marshakov1999} and references therein) which can now be characterised by a quantum curve \cite{Mironov2010a}. And this curve is, actually, equivalent to a TQ-relation upon a (quantum) change of variables \cite{Bourgine2012a}, while the new integrable system is the bi-spectral dual of the previous one \cite{Zenkevich2011}. Of course, of great interest for integrable model theory is the TQ-relation, which can also be obtained directly from the Nekrasov partition function by extremising the sum over Young diagrams \cite{Poghossian2010}, a technique that goes back to \cite{Nekrasov2003a}. Moreover, another key feature of integrability has come out of the blue since the original paper \cite{Nekrasov2009} in the form of a non-linear integral equation (NLIE) in the complex plane \cite{nlie} resembling a Thermodynamical Bethe Ansatz (TBA) equation \cite{Zam-1}. Subsequently, this equation was derived in full detail by \cite{Meneghelli2013,Bourgine2014} upon using the Mayer cluster expansion technique \cite{Mayer1940,Mayer1941}. 

%The solution of this NLIE coincide with the counting function of the integrable systems characterized by the TQ-relation previously mentioned \cite{Bourgine2015b}.

In the present manuscript, we go beyond this limit and present an explicit formula for the first $\e_2$-correction to the prepotential. This result should pave the way to a better understanding of the full $\e_2$-deformation and its meaning as quantum integrable system characterised by a TBA/NLIE. This should ultimately lead to a richer algebraic structure similar to the SHc Hopf algebra as we know how, along the other way around, the NS limit of the SHc algebra can be obtained \cite{Bourgine2014a}. For convenience, we focus on $\mathcal{N}=2$ SYM with a single $SU(N_c)$ gauge group and a number of fundamental flavours. However, we believe that our mathematical construction and results be easily generalisable to arbitrary quivers (along the lines of \cite{Bourgine2014a}).

In fact, with respect to the latter reference (exploiting the methods of Mayer cluster expansion) we shall add an efficient field theory argument. The latter is firstly used to disentangle the long- and short-range interactions between instantons: this is performed in section $2$. Then, we treat the short-range interactions by the Mayer cluster expansion technique in section $3$ and achieve formul{\ae} (\ref{short_1}) and (\ref{Fshort}) concerning the (short-range) prepotential \footnote{This part is rather technical and its details could be skipped at first reading.}. Eventually, the two arguments are combined in section $4$ to build up our proposal for the prepotential at the order $O(\e_2)$. In the last section, this proposal is checked in the simple case of pure $U(1)$ SYM up to the order $O(\L^4)$. A more detailed computation involving only the combinatorics of the Mayer expansion is to appear shortly \cite{Bourgine2015a}.

\section{Nekrasov partition function as Gaussian correlation of exponential fields}
In the expression derived by N. Nekrasov in \cite{Nekrasov2003}, the gauge coupling expansion of the instanton partition function for $\mathcal{N}=2$ $SU(N_c)$ SQCD takes the form of a series over coupled integrals
\begin{equation}\label{PF}
\CZ=\sum_{N=0}^\infty \dfrac{\Lambda^N}{N!}\left(\dfrac{\e_+}{\e_1\e_2}\right)^N\int{\prod_{i=1}^NQ(\phi_i)\dfrac{d\phi_i}{2i\pi}\prod_{\superp{i,j=1}{i<j}}^NK(\phi_i-\phi_j)} \, ,
\end{equation}
where the $\O$-background equivariant deformation parameters $\e_1,\e_2$ are supposed to have a positive imaginary part, $\Lambda$ is the dynamical scale and $\e_+=\e_1+\e_2$. In formula (\ref{PF}) and then in the following, integrations are performed along a contour that surrounds the upper half-plane, including the real axis but excluding possible singularities at the complex infinity. The potential $Q(x)$ is a rational function that can be expressed as the ratio of matter and gauge polynomials,
\begin{equation}
Q(x)=\dfrac{\prod_{f=1}^{N_f}(x-m_f)}{\prod_{l=1}^{N_c}(x-a_l)(x+\e_+-a_l)},
\end{equation}
where $N_f$ and $N_c$ denote respectively the number of flavors (fundamental massive hypermultiplets) and colors (adjoint gauge multiplets). The coulomb branch vacuum expectation values (vevs) are assumed to be real but inside the integration contour, i.e. $a_l\in\mathbb{R}+i0$. The universal kernel for $\mathcal{N}=2$ SYM with a single gauge group ($A_1$ quiver) is 
\begin{equation}
K(x)=\dfrac{x^2(x^2-\e_+^2)}{(x^2-\e_1^2)(x^2-\e_2^2)}.
\end{equation}
Since we consider nested integrals, we need a prescription for the poles at $\phi_i=\phi_j\pm\e_1$ and $\phi_i=\phi_j\pm\e_2$: we assume that the variable $\phi_j$ is real, taking the poles at $\phi_i=\phi_j+\e_1$ and $\phi_i=\phi_j+\e_2$, but not their counterparts at $\phi_i=\phi_j-\e_1$ and $\phi_i=\phi_j-\e_2$. For other gauge groups, the prescription is more subtle and has been discussed in \cite{Hwang2014}.

In order to exploit quantum field theory methods, we factorise the kernel into two pieces
\begin{equation}
K(x)=(1+\e_2 p(x))e^{\e_2 k(x)},\quad\text{with}\quad p(x)=\dfrac{\e_2}{x^2-\e_2^2},\quad k(x)=\e_2^{-1}\log\left(1-\dfrac{2\e_1+\e_2}{x^2-\e_1^2}\e_2\right)\, ,
\end{equation}
where the 'universal' kernel $p(x)$ contains singularities at $x=\pm\e_2$ that pinch the integration contour only in the NS limit $\e_2\to0$. It can be interpreted as a strong short-range interaction responsible for the formation of bound states of instantons \cite{Meneghelli2013,Bourgine2014}. On the contrary, the long-range interaction $k(x)$ is free of such singularities and can be treated by standard Mayer expansion methods \cite{Bourgine2013}. Note that the logarithmic branch cut is not relevant in our context since at a finite order in $\e_2$, the kernel $k(x)$ is a rational function of $x$ of order $O(1)$.

Now, long- and short-range interactions can be separated by the introduction of a Gaussian field $X(x)$ with propagator
\begin{equation}
\la X(x) X(y)\ra = \e_2 k(x-y).
\end{equation}
The corresponding action is a kinetic term with a kernel being the inverse of the propagator,
\begin{equation}
\CS_\text{long}[X]=\dfrac1{2\e_2}\int{\dfrac{dxdy}{(2i\pi)^2}t(x-y)X(x)X(y)},\quad \int{t(x-z)k(z-y)\dfrac{dz}{2i\pi}}=2i\pi\d(x-y),
\end{equation}
and any correlator \, $\cdots$ \,  is understood as Gaussian functional integral average,
\begin{equation}
\la\cdots\ra=\dfrac1{\CN_\text{long}}\int{D X e^{- \CS_\text{long}[X]}\ \cdots\ },\qquad \CN_\text{long}=\int{D X e^{- \CS_\text{long}[X]}}.
\end{equation}
As a consequence of the Wick theorem or Gaussian integration\footnote{Alternatively, this identity can be obtained by simple Gaussian integration of the partition function with a source term 
\begin{equation}
\CZ_\text{long}[J(x)]=\int{D X e^{- \CS_\text{long}[X]+\int{J(x)X(x)dx}}}=\CZ_\text{long}[0]e^{\frac{\e_2}{2}\int{dxdy J(x)J(y)k(x-y)}}.
\label{expXJ}
\end{equation}
For this formula is the infinite dimensional $d\rightarrow \infty$ version of
\begin{equation}
\langle e^{X_1 J_1} e^{X_2 J_2} \cdots e^{X_d J_d} \rangle =\sqrt{\textrm{det}T} \int \prod_{i=1}^d \frac{d X_i}{\sqrt{2\pi}} e^{-\frac{1}{2} \sum\limits _{i,j=1}^d X_i T_{ij} X_j}  e^{ \sum\limits_{i=1}^d X_i J_i}= e^{\frac{1}{2} \sum\limits_{i,j=1}^d J_i G_{ij} J_j}   \, ,
\label{expXJ-d}
\end{equation}
with propagator $G=T^{-1}$, where we choose, for the continuum limit of the external field $J_i\rightarrow J(x)$, the configuration of point-like sources $J(x)=\sum_{i=1}^N  \delta(x-\phi_i)$, {\it cf.} \cite{Fioravanti2015} for the connected application to gluon scattering amplitudes ({\it cf.} also the last section, Perspectives).},

\begin{equation}
\la\prod_{i=1}^N e^{X(\phi_i)}\ra=e^{\frac12 k(0)\e_2 N}\prod_{\superp{i,j=1}{i<j}}^{N}e^{\e_2 k(\phi_i-\phi_j)}.
\end{equation}
This property allows us to perform an Hubbard-Stratonovich transformation on the instanton partition function (\ref{PF}) \begin{equation}\label{Zshort}
\CZ=\la \CZ_\text{short}[X]\ra,\quad\text{with}\quad \CZ_\text{short}[X]=\sum_{N=0}^\infty \dfrac{q^N\e_2^{-N}}{N!}\int{\prod_{i=1}^{N}Q(\phi_i)e^{X(\phi_i)}\dfrac{d\phi_i}{2i\pi}\ \prod_{\superp{i,j=1}{i<j}}^{N}\left(1+\e_2p(\phi_{ij})\right)},
\end{equation}
where we introduced the (finitely) renormalised gauge coupling $q=e^{-\frac12 k(0)\e_2}\Lambda\e_+/\e_1$ and the shortcut notation $\phi_{ij}=\phi_i-\phi_j$. Before performing the $\e_2$-expansion of the quantity $\CZ_\text{short}[X]$ in the next section, we would like to make a remark. It is possible to introduce a second field $\chi(x)$ with Gaussian action
\begin{equation}
\CS_\text{short}[\chi]= \dfrac1{2\e_2}\int{\dfrac{dxdy}{(2i\pi)^2}\t(x-y)\chi(x)\chi(y)},
\end{equation}
and the kinetic kernel being the inverse of the logarithm of the $p$-propagator,
\begin{equation}
\int{\t(x-z)\k(z-y)\dfrac{dz}{2i\pi}}=2i\pi\d(x-y),\quad \k(x)=\e_2^{-1}\log\left(1+\e_2p(x)\right).
\end{equation}
The Nekrasov partition function takes the form of a Gaussian average with respect to the two independent fields,
\begin{equation}
\CZ=\la\la \exp\left(\dfrac{q}{\e_2}\int{Q(\phi)e^{X(\phi)}e^{\chi(\phi)}\dfrac{d\phi}{2i\pi}}\right)\ra\ra.
\end{equation}
In this expression the usual Wick regularisation ({\it e.g.} no equal times contractions) will be needed for the logarithmic divergence of $\k(x)$ in $x=0$, as for the Coulomb gas case of 2D conformal field theories. Instead, without pinching, {\it i.e.} $p\equiv 0$, then $\chi\equiv 0$ and the (combinatorial) results by \cite{Bourgine2013} can be easily obtained by saddle point method ({\it cf.} also below). This simple fact gave us inspiration for the following methodology.

\section{Treatment of the short range interaction}
In order to derive the expression of $\CZ_\text{short}[X]$ at subleading order in $\e_2$, we will treat the quantity $U(x)=qQ(x)e^{X(x)}$ as a meromorphic potential. This treatment is at the moment not fully justified since $X(x)$ is actually a quantum field, but the result we obtain coincide with an alternative, rigorous derivation by performing the Mayer cluster expansion of the original partition function $\CZ$ \cite{Bourgine2015a}.

The Mayer expansion \cite{Mayer1940,Mayer1941} of the short-range prepotential $\CFs[X]=\e_2\log \CZ_\text{short}[X]$ is a sum over connected clusters $\D_l$ with $l$ vertices in which two vertices are connected by at most one link. To each vertex $i\in V(\D_l)$ is associated the integration measure $U(\phi_i)d\phi_i/2i\pi$, and to each link $<ij>\in E(\D_l)$ the short range interaction $\e_2p(\phi_i-\phi_j)$,
\begin{equation}
\CFs[X]=\sum_{l=1}^\infty\sum_{\D_l}\dfrac{\e_2^{-(l-1)}}{\s(\D_l)}\int{\prod_{i\in V(\D_l)}U(\phi_i)\dfrac{d\phi_i}{2i\pi}\prod_{<ij>\in E(\D_l)}\e_2p(\phi_i-\phi_j)}.
\label{free-energy}
\end{equation}
The symmetry factor $\s(\D_l)$ is the cardinal of the group of automorphisms of the cluster $\D_l$. It is also useful to introduce the dressed vertex $\Ys(x)$ obtained as the functional derivative
\begin{equation}\label{Ys_CFs}
\Ys(x)=2i\pi U(x)\dfrac{\d \CFs}{\d U(x)}.
\end{equation}
This quantity expands as a sum over rooted clusters $\D_l^x$ with root $x$,
\begin{equation}
\Ys(x)=U(x)\sum_{l=1}^\infty\sum_{\D_l^x}\dfrac{\e_2^{-(l-1)}}{\s(\D_l^x)}\int{\prod_{\superp{i\in V(\D_l^x)}{i\neq x}}U(\phi_i)\dfrac{d\phi_i}{2i\pi}\prod_{<ij>\in E(\D_l^x)}\e_2p(\phi_i-\phi_j)}.
\end{equation}
There are two main differences with respect to the prepotential (\ref{free-energy}) both due to the fact that we are dealing with rooted clusters: we do not integrate on the root variable $x$, and $\s(\D_l^x) \leq \s(\D_l)$ as, of course, the automorphisms of a rooted cluster keep the root fixed. The short-range free energy $\CFs[X]$ can also be expanded over rooted clusters due to the property (B.3) of \cite{Bourgine2013},
\begin{equation}
\CFs[X]=\int{U(x)\dfrac{dx}{2i\pi}\sum_{l=1}^\infty\dfrac1l\sum_{\D_l^x}\dfrac{\e_2^{-(l-1)}}{\s(\D_l^x)}\int{\prod_{\superp{i\in V(\D_l^x)}{i\neq x}}U(\phi_i)\dfrac{d\phi_i}{2i\pi}\prod_{<ij>\in E(\D_l^x)}\e_2p(\phi_i-\phi_j)}}.
\end{equation}
It was shown in \cite{Bourgine2014} that at first order in $\e_2$ the potentials can be transferred to the root of the clusters $\D_l^x$, and the remaining integral evaluated exactly to give \footnote{The integral
\begin{equation}
I_l=\e_2^{-(l-1)}\sum_{\D_l^x}\dfrac1{\s(\D_l^x)}\int{\prod_{i=1}^{l-1}\dfrac{d\phi_i}{2i\pi}\prod_{<ij>\in E(\D_l^x)}\e_2p(\phi_{ij})}=\dfrac1l
\end{equation}
has been computed in \cite{Bourgine2014}.}
\begin{align}
\begin{split}\label{short_1}
\CFs^{(0)}&=\sum_{l=1}^\infty \int{\dfrac1{l}I_l\ U(x)^l\dfrac{dx}{2i\pi}}=\int{\text{Li}_2(U(x))\dfrac{dx}{2i\pi}},\\
\Ys^{(0)}(x)&=\sum_{l=1}^\infty I_l \ U(x)^l=-\log(1-U(x)).
\end{split}
\end{align}
The first of these gives rise to the Nekrasov-Shatasvili action (and thus prepotential as $\e_2\rightarrow 0$) once inserted in the first of (\ref{Zshort}). Here we want to go beyond this order and give a proposal, which has a little conjectural basis, for the sub-leading $\e_2$ order. A naive approach would consists in employing a linear approximation for the potential, 
\begin{equation}\label{YG_lin}
U(\phi_i)\simeq U(x)+(\phi_i-x)U'(x)+O(\e_2^2).
\end{equation}
However, this does not work because it overlooks the contributions of the poles of the potential. For instance, the contribution of a rooted cluster with a single $p$-link and two vertices reads
\begin{equation}
U(x)\int{p(x-y)U(y)\dfrac{dy}{2i\pi}}=\hf U(x)U(x)^+=\hf U(x)\left[U(x)+\e_2\nabla U(x)+O(\e_2^2)\right],
\end{equation}
where, for any meromorphic function $U(x)$, we define $U(x)^+=U_\text{reg.}(x+\e_2)+U_\text{sing.}(x-\e_2)$ or at infinitesimal level the operator $\nabla$
\begin{equation}\label{def_nabla}
\nabla U(x)=U_\text{reg.}'(x)-U_\text{sing.}'(x) \, ,
\end{equation}
by means of the decomposition $U(x)=U_\text{reg.}(x)+U_\text{sing.}(x)$ with $U_\text{reg.}(x)$ analytic inside the contour of integration while $U_\text{sing.}(x)$ analytic outside. If $U(x)$ had no singularities inside the integration contour, we could replace the symbol $\nabla$ with a partial derivative $\p_x$ and the result would be the same as in the linear approximation (\ref{YG_lin}). In this case, a formula for $\Ys(x)$ exact in $\e_2$ has been derived in \cite{Bettelheim2014} based on an earlier formula obtained by Moore, Nekrasov and Shatashvili in \cite{Moore1997}. Unfortunately, this case is not relevant to the computation of $\CZ_\text{short}[X]$ and we present here an alternative method.

\subsection{Articulation links and irreducible clusters}
In the clusters $\D_l$ it is important to distinguish two types of links. An \textbf{articulation link} of a connected cluster is a link whose removal break the cluster into two disconnected pieces \cite{Andersen1977}. The cluster is said to be ($1$ particle) {\bf reducible}. Note that if the cluster is a tree, all its links are articulation links. A cluster without any articulation link is said to be ($1$ particle) \textbf{irreducible}. An example of irreducible clusters is the necklace, i.e. a ring of $l$ vertices and $l$ links. In \cite{Bourgine2015a}, we have computed the contributions of all necklaces and observed that they do not exhibit any corrections of order $O(\e_2)$. We have further computed the irreducible clusters with four vertices for a potential corresponding to $U(1)$ SYM and we have found the same conclusion. It led us to conjecture that irreducible clusters do not provide any correction at the subleading order (although higher order corrections are present).

We need a stronger form of this conjecture. The $\e_2$-corrections we are trying to compute are associated to the links of the cluster. Heuristically, the total $\e_2$-correction is the sum of the corrections brought by each link and due to a sort of linearisation. Links of a generic cluster can be either articulation links or not. We claim that only articulation links bring non-zero subleading $\e_2$-corrections. To justify this claim, we consider a link of an arbitrary cluster which is not an articulation link, and we compute its contribution to $\CFs$ at the order $O(\e_2)$. We denote the two extremities of the link $x$ and $y$, and $\D_l^{x,y}$ the bi-rooted cluster with $l$ vertices obtained by removing this link. Since we are interested in the $\e_2$-correction of a specific link, it is enough to consider the leading order of the integrals associated to the cluster $\D_l^{x,y}$. At this order, the potential of each vertex can be transferred to a root, say $x$, and produces $U(x)^l$. The remaining integral contribution will be denoted $\G_l(x,y)$, and the integral associated to the original cluster reads
\begin{equation}
\G_l=\int{\dfrac{dxdy}{(2i\pi)^2}p(x-y)U(x)^l\G_l(x,y)},\quad \G_l(x,y)=\e_2^{2-l}\int{\prod_{\superp{i\in V(\D_l^{x,y})}{x\neq i\neq y}}\dfrac{d\phi_i}{2i\pi}\prod_{<ij>\in E(\D_l^{x,y})}\e_2p(\phi_i-\phi_j)}.
\end{equation}
It is useful to rewrite this expression as
\begin{equation}
\G_l=\int{\dfrac{dx}{2i\pi}U(x)^l\nu_l(x,x)},\quad \nu_l(x,z)=\int{\dfrac{d y}{2i\pi}\G_l(x,y)p(y-z)},
\end{equation}
where the function $\nu_l(x,x)$ would contain possible $\e_2$-corrections coming from the $p$-link we singled out.

Due to the invariance under translations of the $\phi_i$, $\G_l(x,y)$ must be a function of the difference $(x-y)$: $\G_l(x,y)=\G_l(x-y)$, and hence $\nu_l(x,z)=\nu_l(x-z)$ as well. Moreover, $\G_l(x)$ is even as can be shown using the sign flip $\phi_i\to -\phi_i$. Convolution with $p$-kernels shift the poles in the upper half plane of $+\e_2$ and of $-\e_2$ in the lower half plane. It turns out that $\G_l(x-y)$ can only have (multiple) poles at $x-y=\pm n\e_2$ with $n=1,2,\dots$. In particular it has no pole at the complex infinity. We deduce the expansion
\begin{equation}
\G_l(x-y)=\sum_{\superp{n=1}{m=0}}^{\infty}\g_{n,m}\left[\dfrac{\e_2^2}{(x-y)^2-n^2\e_2^2}\right]^{m+1}  \ ,
\end{equation}
where due to the scale invariance under $x,y\to \a x,\a y$, $\e_2\to \a\e_2$ and $\phi_i\to\a \phi_i$, the coefficients $\g_{n,m}$ are $\mathbb{C}$-numbers independent of $\e_2$. Their value depends on the explicit form of the cluster $\D_l^{x,y}$, but at fixed $l$ only a finite amount of coefficients are non-zero. Thanks to the formula
\begin{equation}
\left[\dfrac{\e_2^2}{x^2-n^2\e_2^2}\right]^{m+1}=(-2n)^{-(m+1)}\sum_{k=0}^m \left(\superp{m+k}{k}\right) \dfrac{\e_2^{m+1-k}}{(2n)^k}\left[\dfrac1{(x+n\e_2)^{m+1-k}}+\dfrac{(-1)^{m+1-k}}{(x-n\e_2)^{m+1-k}}\right],
\end{equation}
it is possible to separate $\G_l(x-y)$ into regular and singular part with respect to one of its variables, and perform the integration over the $p$-link that has been singled out to obtain $\nu_l(0)$. We observe that $\nu_l(0)$ is a $\mathbb{C}$-number independent of $\e_2$:
\begin{equation}
\nu_l(0)=\sum_{\superp{n=1}{m=0}}^{\infty}\g_{n,m}\ (-1)^{m+1}\sum_{k=0}^m \left(\superp{m+k}{k}\right) \dfrac{1}{\left(2n(n+1)\right)^{m+1-k}}.
\end{equation}
This is a strong hint that $p$-links that are not articulation link do not provide any $\e_2$-correction.

% \begin{equation}
% \nu(x,y)=\hf\sum_{\superp{n=1}{m=0}}^{\infty}\g_{n,m}\ (-2n)^{-(m+1)}\sum_{k=0}^m C_{m+k}^k \dfrac{\e_2^{m+1-k}}{(2n)^k}\left[\dfrac1{(x-y+(n+1)\e_2)^{m+1-k}}+\dfrac{(-1)^{m+1-k}}{(x-y-(n+1)\e_2)^{m+1-k}}\right].
% \end{equation}

\subsection{Considering only trees}
To obtain a better insight on the contributions of articulation links, we first restrict the summation over clusters with a tree structure $\D_{T,l}^x$ and examine the corresponding (tree) dressed vertex
\begin{equation}
\YsT(x)=U(x)\sum_{l=1}^\infty\sum_{\D_{T,l}^x}\dfrac{\e_2^{-(l-1)}}{\s(\D_{T,l}^x)}\int{\prod_{\superp{i\in V(\D_{T,l}^x)}{i\neq x}}U(\phi_i)\dfrac{d\phi_i}{2i\pi}\prod_{<ij>\in E(\D_{T,l}^x)}\e_2p(\phi_i-\phi_j)}.
\end{equation}
Due to the restriction upon tree clusters, this function obeys the following functional equation \cite{Bourgine2014},
\begin{equation}
\YsT(x)=U(x)\exp\left(\int{p(x-y)\YsT(y)\dfrac{dy}{2i\pi}}\right)=U(x)e^{\frac12 \YsT(x)^+},
\end{equation}
the second equality being obtained by an explicit computation of the convolution with the kernel $p$. Expanding in $\e_2$ with the potential $U(x)$ considered finite, we find
\begin{equation}
\YsT(x)=\YsT^{(0)}(x)+\e_2\YsT^{(1)}(x)+O(\e_2^2),\quad \YsT^{(0)}(x)=U(x)e^{\frac12 \YsT^{(0)}(x)},\quad \YsT^{(1)}(x)=\dfrac{\YsT^{(0)}(x)}{2-\YsT^{(0)}(x)}\nabla \YsT^{(0)}(x),
\label{0-1-rooted}
\end{equation}
with the operator $\nabla$ defined in (\ref{def_nabla}). At first order, the generating function can also be expressed in terms of the tree function $T(z)$ \footnote{The tree function is related to the Lambert $W$ function on the principal branch as $T(z)=-W(-z)$. It satisfies
\begin{equation}
z=T(z)e^{-T(z)},\quad T(z)=\sum_{n=1}^\infty{\dfrac{z^n n^n}{n\times n!}},\quad zT'(z)=\dfrac{T(z)}{1-T(z)}.
\end{equation}}
\begin{equation}
\YsT^{(0)}(x)=2T(U(x)/2).
\end{equation}
It is enlightening to recover this result from a direct computation. In fact, we have the property that the potential at each vertex can be transferred to the root, giving
\begin{equation}
\YsT^{(0)}(x)=\sum_{l=1}^\infty U(x)^l\sum_{\D_{T,l}^x}\dfrac{\e_2^{-(l-1)}}{\s(\D_{T,l}^x)}\int{\prod_{\superp{i\in V(\D_{T,l}^x)}{i\neq x}}\dfrac{d\phi_i}{2i\pi}\prod_{<ij>\in E(\D_{T,l}^x)}\e_2p(\phi_i-\phi_j)}.
\end{equation}
As shown in \cite{Bourgine2014}, the remaining integral depends only on the number of links and not on the explicit form of the tree. It evaluates to $(\e_2/2)^{l-1}$. The remaining sum over rooted trees with the appropriate symmetry factors can be evaluated using the Cayleigh formula, as done again in \cite{Bourgine2014} (formula B.1). We obtain
\begin{equation}\label{Taylor_YsT0}
\YsT^{(0)}(x)=2\sum_{l=1}^\infty\left(\dfrac{U(x)}{2}\right)^l\sum_{\D_{T,l}^x}\dfrac1{\s(\D_{T,l}^x)}=2\sum_{l=1}^\infty\left(\dfrac{U(x)}{2}\right)^l\dfrac{l^{l-1}}{l!},
\end{equation}
which is indeed the Taylor expansion of the tree function.

\begin{figure}[!t]
\centering
\includegraphics[width=5cm]{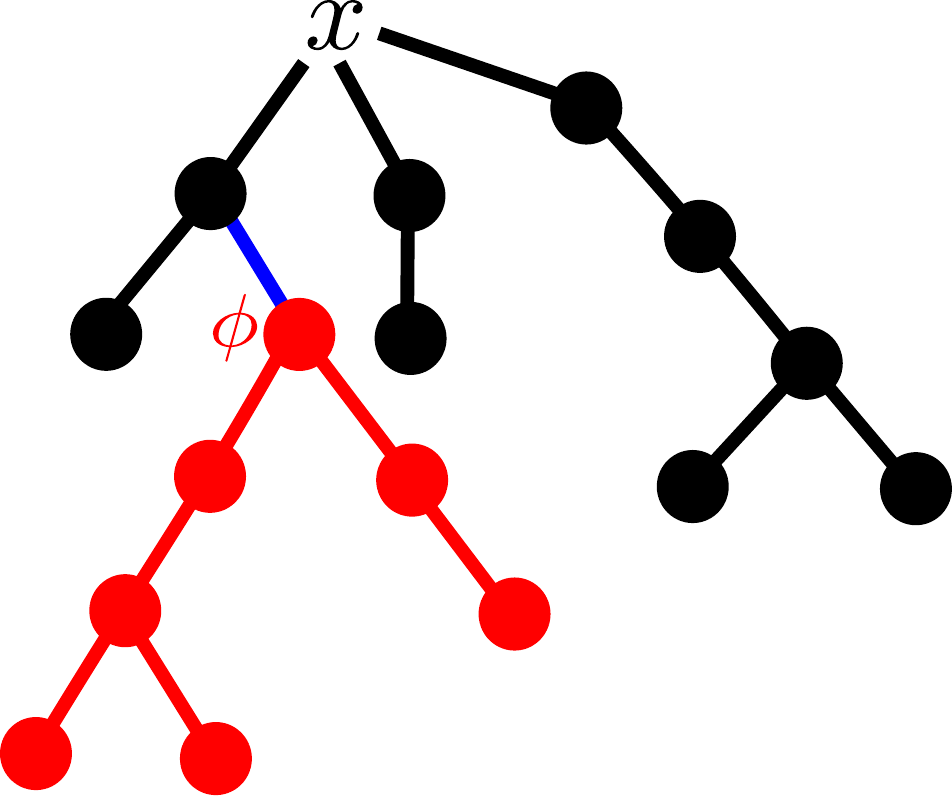}
\caption{A rooted tree decomposed into two rooted subtrees (black and red) by choosing a link (in blue).}
\label{subtree}
\end{figure}

Actually, this direct derivation of the leading order provides us a nice interpretation of next-to-leading contribution, the last of (\ref{0-1-rooted})
\begin{equation}
\YsT^{(1)}(x)=\dfrac{T(U(x)/2)}{1-T(U(x)/2)}\nabla \YsT^{(0)}(x)=\hf U(x) T'(U(x)/2)\nabla \YsT^{(0)}(x)=\sum_{l=1}^\infty\left(\dfrac{U(x)}{2}\right)^l\dfrac{l^l}{l!}\nabla \YsT^{(0)}(x) \, ,
\end{equation}
where, for the last equality, we have used the above Taylor expansion (of the derivative) of the tree function. In fact, following back in the reverse order the steps to the derivation of the formula (\ref{Taylor_YsT0}), we can engineer a cluster expansion formula for $\YsT^{(1)}(x)$,
\begin{equation}
\YsT^{(1)}(x)\simeq\hf U(x)\sum_{l=1}^\infty\e_2^{-(l-1)}\sum_{\D_{T,l}^x}\dfrac{l}{\s(\D_{T,l}^x)}\int{\prod_{\superp{i\in V(\D_{T,l}^x)}{i\neq x}}U(\phi_i)\dfrac{d\phi_i}{2i\pi}\prod_{<ij>\in E(\D_{T,l}^x)}\e_2p(\phi_i-\phi_j)} \nabla \YsT^{(0)}(x),
\end{equation}
where we denoted $A\simeq B$ for $A=B+O(\e_2)$. The extra symmetry factor $l$ corresponds to the possibilities to graft a leaf to the tree $\D_{T,l}^x$ on any of its vertices, i.e.
\small
\begin{equation}\label{CYT1}
\YsT^{(1)}(x)\simeq U(x)\sum_{l=1}^\infty\e_2^{-(l-1)}\sum_{\D_{T,l}^x}\dfrac{1}{\s(\D_{T,l}^x)}\int{\!\!\!\prod_{\superp{i\in V(\D_{T,l}^x)}{i\neq x}}\!\!\!\! U(\phi_i)\dfrac{d\phi_i}{2i\pi}\!\!\!\!\prod_{<ij>\in E(\D_{T,l}^x)}\!\!\!\!\!\!\!\!\e_2p(\phi_i-\phi_j)} \int{\!\!\!\!\sum_{\a\in V(\D_{T,l}^x)}\!\!\!\!p(\phi-\phi_\a)\ \nabla \YsT^{(0)}(\phi)\dfrac{d\phi}{2i\pi}}.
\end{equation}
\normalsize
In this formula, the extra vertex $\phi$ is grafted to the tree at the vertex $\phi_\a$ via a $p$-link. This vertex bear the potential $\nabla \YsT^{(0)}(\phi)$ instead of $U(\phi)$. But our formula can also be obtained in another way: consider a tree $\D_{T,l}^x$ and select a link that will later correspond to $p(\phi-\phi_\a)$. Removing this link, one obtains two clusters, one which is rooted by $x$ called $T_1^x$ (in black in figure \refOld{subtree}), and a subtree of the original tree denoted $T_2^\phi$ with a new root $\phi$ where the link we removed used to end (in red on figure \refOld{subtree}). To compute the $\e_2$-correction to this specific $p$-link, we can approximate at first order each of the two clusters and the generating function associated to $T_2$ is $\YsT^{(0)}(\phi)$. The correction from the $p$-link in blue appears in the integral
\begin{equation}
\int{p(\phi_\a-\phi)\YsT^{(0)}(\phi)\dfrac{d\phi}{2i\pi}}=\hf\YsT^{(0)}(\phi_\a)^+=\hf\YsT^{(0)}(\phi_\a)+\hf\e_2\nabla\YsT^{(0)}(\phi_\a)+O(\e_2^2),
\end{equation}
where $\phi_\a$ is any vertex of $T_1^x$. The second term in the RHS is interpreted as the subleading $\e_2$-correction brought by the blue $p$-link. It only remains to compute at first order the integral associated to the vertices of $T_1^x$ which exactly corresponds to the formula (\ref{CYT1}).

\subsection{Contributions from articulation links}
The $\e_2$-corrections provided by the articulation links are similar to those obtained in (\ref{CYT1}) for the trees. The only difference is that now the articulation link splits the cluster into two generic clusters instead of two trees. The correction $\Ys^{(1)}(x)$ is thus obtained by grafting to a rooted cluster a leaf with the potential $\nabla \Ys^{(0)}$ that generates the second cluster:
\begin{equation}
\Ys^{(1)}(x)\simeq\hf U(x)\sum_{l=1}^\infty\e_2^{-(l-1)}\sum_{\D_l^x}\dfrac{l}{\s(\D_l^x)}\int{\prod_{\superp{i\in V(\D_l^x)}{i\neq x}}U(\phi_i)\dfrac{d\phi_i}{2i\pi}\prod_{<ij>\in E(\D_l^x)}p(\phi_i-\phi_j)} \nabla \Ys^{(0)}(x) \, ,
\end{equation}
upon recalling (\ref{short_1}) and $U(x)=qQ(x)e^{X(x)}$. Transferring the potential to the root, we find (by also using $I_l=1/l$ and (\ref{short_1}))
\begin{equation}
\Ys^{(1)}(x)=\hf \sum_{l=1}^\infty l I_l U(x)^l  \nabla \Ys^{(0)}(x)=\hf\dfrac{U(x)}{1-U(x)}\nabla \Ys^{(0)}(x)=\hf \left(e^{\Ys^{(0)}(x)}-1\right)\nabla \Ys^{(0)}(x).
\end{equation}
The correction to the free energy $\CFs$ presents an extra factor ($1/l$ and)  $1/2$ due to the possibility to exchange the two sub-clusters in absence of a root
\begin{equation}\label{Fshort}
\CFs^{(1)}=\dfrac14\int{\dfrac{dx}{2i\pi} U(x)\sum_{l=1}^\infty\e_2^{-(l-1)}\sum_{\D_l^x}\dfrac{1}{\s(\D_l^x)}\int{\prod_{\superp{i\in V(\D_l^x)}{i\neq x}}U(\phi_i)\dfrac{d\phi_i}{2i\pi}\prod_{<ij>\in E(\D_l^x)}p(\phi_i-\phi_j)} \nabla \Ys^{(0)}(x)}.
\end{equation}
By factorisation of the potential, we recover the integral $I_l$, which, eventually, leads us to the correction
\begin{equation}
\CFs^{(1)}[X]=\dfrac14\int{\dfrac{dx}{2i\pi} \Ys^{(0)}(x)\nabla \Ys^{(0)}(x)} \, .
\end{equation}

\section{Prepotential at subleading order}
Upon plugging the previous result for $\CZ_\text{short}$ into the first of (\ref{Zshort}), we can give an explicit path integral expression for the Nekrasov partition function \footnote{In principle this procedure could be carried on at any order in $\e_2$.} 
\begin{equation}
\CZ\simeq\la\exp\left(\dfrac1{\e_2}\int{\text{Li}_2\left(qQ(x)e^{X(x)}\right)\dfrac{dx}{2i\pi}}+\dfrac14\int{\log\left(1-qQ(x)e^{X(x)}\right)\nabla\log\left(1-qQ(x)e^{X(x)}\right)\dfrac{dx}{2i\pi}}\right)\ra \, ,
\end{equation}
where the term involving the operator $\nabla$ (whose action on the logarithm is well-defined only perturbatively in $q$, at any finite order) yields the exact subleading correction.

In the previous expression we may re-introduce the initial kernel $k(x-y)$ in place of its inverse $t(x-y)$ via a ghost field $\rho(x)$ (Hubbard-Stratonovich transform) \footnote{Although convenient, the introduction of the extra field $\rho(x)$ is not strictly necessary. It is indeed possible to perform directly a semi-classical treatment of the quantum field $X$. The equations of motion associated to the variation of $X$ reads, at leading order,
%\begin{equation}
%2i\pi\dfrac{\d\CS[X]}{\d X(x)}=-\int{t(x-y)X(y)\dfrac{dy}{2i\pi}}-\log\left(1-qQ(x)e^{X(x)}\right).
%\end{equation}
upon performing a convolution with the original kernel $k(x)$, as the following TBA-NLIE
\begin{equation}
X(x)+\int{k(x-y)\log\left(1-qQ(y)e^{X(y)}\right)\dfrac{dy}{2i\pi}}=0.
\end{equation}
Deriving once more, the Hessian kernel reads
\begin{equation}
\dfrac{\d^2 \CS[X]}{\d X(x)\d X(y)}= \frac{1}{(2i\pi)^2} t(x-y) - \frac{1}{2i\pi} \d(x-y)\dfrac{qQ(x)e^{X(x)}}{1-qQ(x)e^{X(x)}} \, .
\label{hessian-1}
\end{equation}
Once convoluted with the kernel $k(x-y)$ this gives the same as the Hessian kernel with two fields $\rho$ and $\vphi$ as computed below.
}; moreover, to better see the correction to the Nekrasov-Shatashvili action (which indeed contains two fields), we shall perform a change of variable $\vphi(x)=-X(x)$, so that
\begin{align}
\begin{split}
\CZ\simeq\int D \rho D \vphi \exp\Bigg(&\dfrac1{2\e_2}\int{k(x-y)\rho(x)\rho(y)dxdy}+\dfrac1{\e_2}\int{\rho(x)\vphi(x)dx}+\dfrac1{\e_2}\int{\text{Li}_2\left(qQ(x)e^{-\vphi(x)}\right)\dfrac{dx}{2i\pi}}\\
&+\dfrac14\int{\log\left(1-qQ(x)e^{-\vphi(x)}\right)\nabla\log\left(1-qQ(x)e^{-\vphi(x)}\right)\dfrac{dx}{2i\pi}}\Bigg).
\end{split}
\end{align}
In the r.h.s., the term of the second line is subleading in $\e_2$ and will be neglected in the derivation of the equations of motion. The remaining term defines the action $\CS_0[\rho,\vphi]$ that gives    
\footnote{At this order, it is not necessary to compute the variation of the term involving the operator $\nabla$. %However, it is interesting to note that this variation can be obtained using the combinatorial relation \ref{Ys_CFs} between $\CFs^{(1)}$ and $\Ys^{(1)}(x)$. It gives the prescription
%\begin{equation}
%\dfrac{\d\nabla \Ys^{(0)}(y)}{\d U(x)}=\d(x-y) \dfrac{e^{\Ys^{(0)}(x)}}{\Ys^{(0)}(x)},\quad\text{or}\quad \dfrac{\d\nabla \rho(y)}{\d \rho(x)}=\dfrac{2i\pi}{\rho(x)}\d(x-y).
%\end{equation}
}
\begin{align}
\begin{split}
&\dfrac{\d \CS_0[\rho,\vphi]}{\d\rho(x)}=-\int{k(x-y)\rho(y)dy}-\vphi(x)=0,\\
&\dfrac{\d \CS_0[\rho,\vphi]}{\d\vphi(x)}=-\rho(x)-\dfrac1{2i\pi}\log\left(1-qQ(x)e^{-\vphi(x)}\right)=0.
\end{split}
\end{align}
These equations allows to determine the classical fields by solving a TBA-NLIE,
\begin{equation}\label{NLIE_phi}
\vphi(x)=\int{k(x-y)\log\left(1-qQ(y)e^{-\vphi(y)}\right)\dfrac{dy}{2i\pi}},\quad 2i\pi\rho(x)=-\log\left(1-qQ(x)e^{-\vphi(x)}\right).
\end{equation}
Once the classical fields are know, {\it i.e.} on-shell, we can compute (minus) the Hessian kernel defined by
\begin{equation}
H(x,y)= - \int{dz\left[\dfrac{\d^2\CS_0}{\d\rho(x)\d\rho(z)}\dfrac{\d^2\CS_0}{\d\vphi(z)\d\vphi(y)}-\dfrac{\d^2\CS_0}{\d\vphi(x)\d\rho(z)}\dfrac{\d^2\CS_0}{\d\rho(z)\d\vphi(y)}\right]},
\end{equation}
and find
\begin{equation}
H(x,y)=\d(x-y)-\dfrac1{2i\pi}k(x-y)\D(y),\quad \D(x)=\dfrac{qQ(x)e^{-\vphi(x)}}{1-qQ(x)e^{-\vphi(x)}}.
\end{equation}
The same result could be obtained from the Hessian kernel with only one field $X$ (\ref{hessian-1}) upon convolution with the kernel $k(x-y)$, which is the contribution to the measure from the introduction of the field $\r$ according to (\ref{expXJ}), (\ref{expXJ-d}). Note that on-shell we also have $\D=e^{2i\pi\rho}-1$. Eventually, taking into account both classical action and one-loop corrections, we find for the prepotential $\CF=\e_2\log\CZ$ in terms of the classical solutions to (\ref{NLIE_phi})
\begin{equation}\label{final}
\CF=\hf\int{\rho(x)\vphi(x)dx}+\int{\text{Li}_2\left(qQ(x)e^{-\vphi(x)}\right)\dfrac{dx}{2i\pi}}+\dfrac{i\pi}2\e_2\int{\rho(x)\nabla\rho(x)dx}-\dfrac{\e_2}{2}\log\det H(x,y)+O(\e_2^2)\, ,
\end{equation}
where, for instance, the logarithm of the determinant of $H(x,y)$ enjoys a Fredholm infinite series formula
\begin{equation}\label{Fredholm}
\log \det H(x,y)=-\sum_{n=1}^\infty\dfrac1{n}\int{\prod_{i=1}^n\D(\phi_i)\dfrac{d\phi_i}{2i\pi}\prod_{i=1}^{n-1}k(\phi_i-\phi_{i+1})\   k(\phi_n-\phi_1)}.
\end{equation}

% \paragraph{Tadpole} The tadpole term (to be properly introduced) can be re-absorbed by a translation of the field $\vphi(x)$. At the order of interest, it has the effect to replace the renormalized coupling $q$ by the bare one $\qb$ in the dilogarithm, and henceforth in the equations of motions that determine the classical fields.

\section{Study of $U(1)$ $\mathcal{N}=2$ SYM}
The Nekrasov instanton partition function of $U(1)$ SYM without matter is particularly simple,
\begin{equation}
\CZ_{U(1)}=\exp\left(\dfrac{\L}{\e_1\e_2}\right).
\label{ZU(1)}
\end{equation}
However, the evaluation of the each term contributing to the prepotential exhibits a non-trivial dependence on the gauge coupling, which is remarkably canceled out order-by-order so to provide the simple result given above. This theory is thus a very good candidate to test our formula (\ref{final}) for the subleading corrections in $\e_2$ to the prepotential. Here we perform this verification up to the order $O(\L^4)$, i.e. four vertices. In particular, this is a good test of our conjecture about the non-contribution of irreducible $p$-clusters at the origin of our formula for $\CZ_\text{short}[X]$.

The potential associated to vertices for pure $U(1)$ SYM reads
\begin{equation}
Q(x)=\dfrac1{(x-a)(x-a+\e_1+\e_2)} \, .
\end{equation}
Note that the Coulomb branch vev $a$ can be set to zero due to the invariance under translations of the instantons position. Furthermore, at the order of interest $q$ and $\L$ can be identified since $q=\L+O(\e_2^3)$. The field $\vphi(x)$ is determined at the order $O(q^3)$ and $O(\e_2)$ after solving by iterations the TBA-NLIE (\ref{NLIE_phi}). Plugging the result in the second equation \label{NLIE} produces $\rho(x)$ at the order $O(q^4)$. Isolating the singularities in the upper half plane at $x=a+n\e_1$ with $n=0,1,2,3$ that contributes to the singular part $\rho_\text{sing.}$, it is possible to compute $\nabla\rho(x)$. From these expressions we deduce the following contributions after evaluating the integrals as sum over residues,
\begin{align}
\begin{split}
&\hf\int{\rho(x)\vphi(x)dx}=-\dfrac{q^2}{2\e_1^3}+\dfrac{q^3}{12\e_1^5}-\dfrac{q^4}{36\e_1^7}+\left(\dfrac{11q^2}{8\e_1^4}-\dfrac{49q^3}{72\e_1^6}+\dfrac{1583q^4}{3456\e_1^8}\right)\e_2+O(q^5,\e_2^2),\\
&\int{\text{Li}_2\left(q Q(x)e^{-\vphi(x)}\right)\dfrac{dx}{2i\pi}}=\dfrac{q}{\e_1}+\dfrac{q^2}{2\e_1^3}-\dfrac{q^3}{12\e_1^5}+\dfrac{q^4}{36\e_1^7}+\left(-\dfrac{q}{\e_1^2}-\dfrac{5q^2}{4\e_1^4}+\dfrac{5q^3}{8\e_1^6}-\dfrac{371q^4}{864\e_1^8}\right)\e_2+O(q^5,\e_2^2),\\
&\dfrac{i\pi}2\e_2\int{\rho(x)\nabla\rho(x)dx}=\left(\dfrac{q^2}{2\e_1^4}-\dfrac{5q^3}{8\e_1^6}+\dfrac{22q^4}{27\e_1^8}\right)\e_2+O(q^5,\e_2^2),\\
&-\dfrac{\e_2}2\log\det H(x,y)=\left(\dfrac{q}{\e_1^2}-\dfrac{5q^2}{8\e_1^4}+\dfrac{49q^3}{72\e_1^6}-\dfrac{2915q^4}{3456\e_1^8}\right)\e_2+O(q^5,\e_2^2).
\end{split}
\end{align}
The determinant of the Hessian kernel is obtained by a consistent truncation of the Fredholm expansion (\ref{Fredholm}) at the order $n=4$. Taking the sum of these terms, we match (\ref{ZU(1)}) with $\CF_{U(1)}=\e_2\log\CZ_{U(1)}=\L/\e_1+O(\L^5,\e_2^2)$.

\section{Perspectives}
We have proposed a formula for the subleading corrections to the prepotential of $SU(N_c)$ $\mathcal{N}=2$ SYM in the Nekrasov-Shatashvili limit. However, two physically justified assumptions are used for the derivation and shall deserve deeper analysis in the near future. First, the field theory argument rest upon the treatment of the exponential of the field as a meromorphic potential. This assumption can be bypassed by a treatment of the full partition function by Mayer expansion. Since this derivation is rather lengthy, we decided to present it elsewhere \cite{Bourgine2015a}. The second point that requires a closer examination, in the close future, is the $\e_2$-expansion of $\CZ_\text{short}$ from which the full partition function can now be derived. 

The results presented here are very general and may find application in numerous problems. For instance, the computation of planar amplitudes in $\mathcal{N}=4$ SYM involves series of nested integrals that are similar to those defining the Nekrasov instanton partition function \cite{BSV}. In particular, they also exhibit at strong coupling a phenomenon where some poles pinch the integration contours, which is responsible for the formation of a heavy 'meson' and its bound states \cite{Fioravanti2015}. Hopefully, the formul{\ae} presented here might be generalised for computing the one-loop correction to classical minimal area result, namely the contribution at order of the inverse string tension $\e_2\sim 1/\sqrt{\l}$ ($\sqrt{\l}$ is also the t'Hooft coupling).

The presence of the TBA-like NLIE at leading order connects the supersymmetric gauge theories to the realm of quantum integrable systems. Turning on the $\e_2$ parameter defines an integrable deformation of these systems that is characterised by the presence of the SHc algebra. However, this deformation is still poorly understood in the perspective of integrable theories. With the derivation of the first order correction, we hope to clarify the situation and find a physical interpretation of the integrable deformation. It is noted that in the Nekrasov-Shatashvili dictionary $\e_2$ plays the role of the inverse length of the thermodynamic limit. It suggests to interpret the $\e_2$-corrections as further finite size corrections to the Yang-Yang functional. It would be interesting to find this interpretation in a simple integrable system, also for the development of the integrability theory.

The possible connections with Painlev\'e sixth equations and isomonodromy problems also deserve further investigation \cite{Troost2013,Litvinov2013}. In the case of an $SU(2)$ gauge group, the NS prepotential is related under the AGT correspondence to the semi-classical conformal block of Liouville field theory. This conformal block can be obtained by solving in the classical limit $b\to0$ the null vector decoupling equation (NVDE) obeyed by a five-points correlator with a degenerate operator of level two inserted. This operation corresponds to insert a surface operator in the AGT dual instanton partition function \cite{Alday2009a}. The NVDE is identified in \cite{Troost2013} with a Schr\"odinger equation obtained in a quantised version of the sixth Painlev\'e equation \cite{Nagoya2009}. Our result should be connected to the $O(\hbar)$ corrections to the WKB solution of this Schr\"odinger problem.

Still in the case of $SU(2)$ SYM, it is possible to take the colliding limit of the potential $Q(x)$ in which the leading order prepotential compute the classical irregular conformal blocks \cite{Gaiotto2009b}. This quantity is related to the accessory parameter of a Mathieu equation (i.e. the energy in the Schr\"odinger form) obtained as the aforementioned classical limit of the NVDE \cite{Piatek2014a}. The formula presented here allows the computation of the first semi-classical correction to the irregular conformal block. This correction could be related to a deformation of the Mathieu equation.

% 
% the quantum Seiberg-Witten curve can be written in the form of a Mathieu equation and it is closely related to the sine-Gordon model \cite{Mironov2010,Maruyoshi2010}.

\medskip
\medskip
{\bf Acknowledgements.}
It is a pleasure to thank A. Lerda and R. Tateo for useful discussions. This project was  partially supported by INFN grant GAST, the UniTo-SanPaolo research  grant Nr TO-Call3-2012-0088, the ESF Network (09-RNP-092 (PESC)) and the MPNS--COST Action MP1210.

\appendix

\bibliographystyle{unsrt}

\end{document}